\documentstyle[amssymb,multicol,aps,prl,floats]{revtex}

\draft

\begin{document}
\title{Small-scale turbulent dynamo}
\author{M. Chertkov$^a$, G. Falkovich$^b$, I. Kolokolov$^{b,c}$ and 
M. Vergassola$^{d}$}
\address{$^{a}$ 
T-13 and  CNLS, Los Alamos National Laboratory, Los Alamos, NM 87545, USA
\\
$^{b}$ Physics of Complex Systems, Weizmann Institute,
Rehovot 76100, Israel\\
$^{c}$ Budker Institute of Nuclear Physics, Novosibirsk 630090, Russia \\
$^{d}$ CNRS, Observatoire de la C\^ote d'Azur, B.P. 4229, 
06304 Nice Cedex 4, France}
\maketitle

\begin{abstract} Kinematic dynamo theory is presented
here for turbulent conductive fluids. We describe how inhomogeneous
magnetic fluctuations are generated below the viscous scale of
turbulence where the spatial smoothness of the velocity permits a
systematic analysis of the Lagrangian path dynamics.  We find
analytically the moments and multi-point correlation functions of the
magnetic field at small yet finite magnetic diffusivity. We show that
the field is concentrated in long narrow strips and describe anomalous
scalings and angular singularities of the multi-point correlation
functions which are manifestations of field's intermittency. The
growth rate of the magnetic field in a typical realization
is found to be half the difference of two
Lyapunov exponents of the same sign.

\end{abstract}

\begin{multicols}{2} It is believed that the magnetic fields of planets,
stars and galaxies have their origin in dynamo action driven by
motions of conducting fluids \cite {M78,P79,CG95}. Inhomogeneous flow
stretches magnetic lines amplifying the field while the field produces
electric currents that dissipate energy and diffuse the field due to
finite resistivity.  The outcome of the competition between
amplification and diffusion depends on the type of the flow.  We
consider the long-standing problem of how turbulence excites
inhomogeneous fluctuations of magnetic field \cite{B50,KN67,K68}.
Since the growth-rate is proportional to velocity gradients, the
fastest growth is for the fluctuations shorter than the viscous scale
of turbulence, they are the first to reach saturation and strongly
influence the subsequent evolution of the system \cite{KA92}.  It is
then important to have a systematic description of the field that has
emerged from the linear dynamo phase.  Consistent description of the
long-time evolution of the small-scale field with the account of
diffusion remained elusive for a long time \cite{CG95}, only the
second moment has been found \cite{KN67,K68}. When diffusivity $%
\kappa $ is small, the field is almost frozen into the fluid and is
expected to grow exponentially like an infinitesimal material line
element.  To what extent this is offset by a transversal contraction
that eventually brings diffusion into play depends on the statistics
of stretching and contraction.  We find below the growth rate of the
field in a typical realization $\gamma=\langle\log B \rangle/t$ for an
arbitrary velocity statistics and derive analytically the whole
function $E_n=\log \langle B^{2n}\rangle /t$ for a short-correlated
velocity. Both $E_n$ and $\gamma$ are finite at $\kappa \to +0$ (what
is called fast dynamo \cite {CG95,VZ72}). Particularly interesting is
that $\gamma=\lambda_1$ in a perfect conductor while
$\gamma\leq\lambda_1/2$ at whatever small resistance, $\lambda_1$
being the largest Lyapunov exponent i.e. the growth rate of a material
line element.

Consider the kinematic stage of dynamo when the only equation
to solve is that for the magnetic field
\begin{equation}
\partial_t {\bf B}+({\bf v}\cdot{\bf \partial}){\bf B}=
({\bf B}\cdot{\bf \partial}){\bf v}+\kappa \triangle {\bf B},   \label{B1}
\end{equation} 
while the velocity statistics is presumed to be known.  In many
astrophysical applications the viscosity-to-diffusivity ratio is large
and there is a wide interval of scales between viscous and diffusive
cut-offs where the velocity is spatially smooth while the magnetic
field has a nontrivial spatial structure to be described below.  Given
the initial condition, the solution of (\ref{B1}) for the smooth case
(when one may substitute ${\bf v}=\hat{\sigma}{\bf r}$ introducing the
local strain $\sigma_{\alpha\beta}=\partial v_\alpha/\partial
r_\beta$) is most conveniently written in Fourier representation
$${\bf B}( {\bf k},t) = \hat{W}\left( t\right) {\bf B} \left({\bf
k}(0),0\right) \exp \left(-\kappa \int_0^t
k^2(t^{\prime})\,dt^{\prime}\right),$$ where the wavevectors evolve as
${\bf k}(t^{\prime})=\hat{W}^T(t,t^{\prime}){\bf k}(t)$ and the final
condition is ${\bf k}(t)={\bf k}$ \cite{MRS85}. The evolution matrix
$\hat{W}$ satisfies $d\hat W(t,t^{\prime})/dt= \hat\sigma(t)\hat
W(t,t^{\prime})$, with $\hat{W}(t',t')={\bf 1}$ and
$\hat{W}(t)=\hat{W}(t,0)$.  We adopt here the methods of Lagrangian
path analysis \cite{94SS,95CFKLa} developed recently for the related
problem of passive scalar \cite{59Bat,74Kra}.  The moments of ${\bf
B}$ are to be calculated by two independent averaging: first,
(trivial) average over the initial statistics and, second, average over
velocity statistics. Without any loss of generality, we assume the
initial statistics to be homogeneous, isotropic and Gaussian, with
zero mean and the variance $\langle B_\alpha({\bf k} ,0)B_\beta({\bf
k'},0)\rangle=P_{\alpha\beta}({\bf k})k^2\,f(k^2)\delta({\bf k}+ {\bf
k'})$.  The solenoidal projector $P_{\alpha\beta}=
\left(\delta_{\alpha\beta}-k_\alpha k_\beta/k^2\right)$ ensures that
${\bf B}$ is divergence-free.  The initial magnetic noise is
concentrated at the scale $L$, we use $f(k^2)=L^5\exp(-k^2L^2)$
whenever an explicit calculation is performed.  Inhomogeneous
advection produces smaller and smaller scales and balances with
diffusion at the scale $r_d= \sqrt{\kappa/\lambda_1}$, the magnetic
Reynolds number $L/r_d$ is assumed to be large.

The wavectors are of order $1/L$ initially and for some period they
remain much smaller than $1/r_d$. This is the stage where the dynamics
is insensitive to diffusion so that the field is frozen into the fluid
like in a perfect conductor, see \cite{CG95,MRS85} and (\ref{moments})
below. At some time $t_d\propto \ln(L/r_d)$, the wavevectors reach
$1/r_d$, transversal contractions bring diffusion into play and the
new regime starts which is the main subject of this paper. It is
supposed that $t_d$ is much larger than the velocity correlation time
$\tau$ and we can then carry over from random matrix theory the
well-known Iwasawa decomposition (see e.g. \cite{GSO87})\,:
$\hat{{W}}(t)={\hat R}{\hat D}{\hat S}$. For any fixed time, $\hat {
R}$ is an ${\rm SO}(3)$ rotation matrix, $\hat{\rm D}$ is diagonal
with ${\rm D}_{ii}(t)=\exp [\rho _{i}(t)]$ and the shearing matrix
$\hat{\rm S}$ is upper triangular with unit elements on the
diagonal. The sum of $\rho_i$'s vanishes by incompressibility and the
ratios $\rho_i/t$ tend at $t\rightarrow\infty$ to the three Lyapunov
exponents $\lambda_i$ (arranged in decreasing order) --- see
e.g. \cite{CPV}.

Since ${\bf {B}}$ is expressed via $\hat{{W}}$, our aim now is to
reformulate the average over $\hat{\sigma}$ into that over
$\hat{{W}}$. The average over $\hat{ R}$ is equivalent to the
integration over the directions of the vectors involved.  The matrix
${\hat S}$ tends for each realization to a time-independent
form. Indeed, it will be shown below that the magnetic field moments
grow exponentially in time and the dominant contributions come from
realizations with $\rho _{1}\gg \rho _{2}$ at $t\to \infty $. For such
realizations the matrix $\hat{\rm S}$ is frozen at large times, the
eigendirections of $\hat{W}^T\,\hat{W}$ do not fluctuate in time and
fix an orthogonal basis\,: ${\bf n}_1$, ${\bf n}_2$, ${\bf n}_3$, with
respective stretching (contraction) rates $2\lambda_1$, $2\lambda_2$,
$2\lambda_3$ \cite{GSO87}. We shall also see below that the
time-independent random matrix elements of $\hat{S}$ influence only
constant factors, not of interest for the space-time dependencies
studied here. The problem is reduced then, on one hand, to the
integration over the angles of the vectors involved and, on the other
hand, to the average over the statistics of $\rho_1$ and $\rho_2$.

The moments of the magnetic field can be
obtained by considering ${\bf B}^2(t)$ averaged over initial
statistics 
$${\bf B}^2(t)=\int {d^3{q}}f(q) e^{-2\kappa {\bf q}\hat\Lambda{\bf
q}} q^2{{\rm Tr}}[\hat{W} \hat{P}({\bf q})\hat{W}^T],$$ then taking
powers and averaging over the velocity.  Here we have made the change of
variables ${\bf q}=\hat{W}^{T}{\bf k}$ that reexpresses the average in
terms of the wavevectors at $t=0$.  When $t\gg |\lambda_3|^{-1}$ the
main contribution to $\hat\Lambda(t)=\int_0^t
dt'\hat{W}^{-1}(t')\hat{W}^{-1,T}(t')$ is given by $t'\gg
|\lambda_3|^{-1}$.  Changing the variables ${\bf q}=\hat{S}^T{\bf Q}$
we eliminate the constant $\hat{S}$-matrix from the diffusive
exponent. The dependence on $\hat{S}$ remains just in the quadratic in
${\bf Q} $ prefactor and in $f$ so that in averaging over the velocity we
may replace $\hat{{S}}$ by the unit matrix. It follows that ${\bf
q}\hat\Lambda {\bf q}=\int_0^t dt \sum Q_{i}^{2}e^{-2\rho _{i}}\equiv
U(Q,\rho)$ and in the ${\bf q}^2{{\rm Tr}}$ term the main contribution
is $e^{2\rho _{1}}(Q_{2}^{2}+Q_{3}^{2})$.  In any given realization
the growth of the field is thus described by a simple formula
$${\bf
B}^{2}(t)\simeq\int {d^{3}{ Q} } f(Q) \exp( -2\kappa U)e^{2\rho
_{1}}(Q_{2}^{2}+Q_{3}^{2})\ .$$  
The dynamics of the field growth is now
clearly seen.  Initially diffusion is unimportant ($\kappa U\ll1$) and
$B^2$ grows as $e^{2\rho_1}$ i.e.  as a square of a material line
element. At $t\simeq t_d=|\lambda_3|^{-1}\log L/r_d$, the diffusive
exponent starts to decrease substantially and the growth rate is
reduced. Asymptotically for $t\gg t_d$, it is clear that the
realizations and the ${\bf q}$'s dominating the growth are such that
the quadratic form $\kappa U$ at the diffusive exponent remain $O(1)$.
Note that for growing functions one has with exponential accuracy
$\int^t dt' \exp(-\rho_3(t'))\propto \exp(-\rho_3(t))$.  The
integration over $Q_3$ is thus restricted within exponentially small
interval: the diffusion exponent remains $O(1)$ only for initial
wavevectors with such a small projection on the contraction direction
${\bf n}_3$ that the respective component does not reach $1/r_d$
during the time $t$ \cite{MRS85}.  Neglecting $Q_{3}^{2}$
comparatively to $Q_2^2$ and omitting numerical factors, we get after
the integration over ${\bf Q}$
\begin{eqnarray}{\bf B}^{2}(t)\simeq \exp[2\rho_1(t)]\{1+(r_d/L)^2
\exp[-2\rho_3(t)]\}^{-1/2}\nonumber\\
\times \left\{1+(r_d/L)^2
\int^t dt' \exp[-2\rho_2(t')]\right\}^{-3/2}.\label{int}
\end{eqnarray}
The first figure bracket reduces for large times to $(L/r_d)\exp\rho_3$, the
geometrical factor due to the orthogonality condition to ${\bf n}_3$. In
the second line the exponential term can be either comparable or
larger than unity depending on the sign of $\rho_2$, 
which corresponds to the two geometrical pictures (cone
{\it vs} pancake in $k$-space) illustrated in \cite{MRS85}.

Moments of (\ref{int}) should be averaged over the probability
distribution ${\cal P}_t(\rho_1,\rho_2)$.  When $t\gg
\lambda_1^{-1},\tau$, the theory of large deviations assures that
${\cal P}_t(\rho_1,\rho_2)$$\propto\exp[-tH(\rho_1/t,\rho_2/t)]$,
where the entropy $H(x,y)$ has a sharp minimum $H=0$ at $x=\lambda_1$,
$y=\lambda_2$ whose width decreases as $t^{-1/2}$ \cite{Ellis}
(for the vector case see, e.g., \cite{BF99}).  
The mean growth rate
$\gamma(t)=\langle \ln B^2(t)\rangle/2t$ is then simply obtained by
taking the logarithm of (\ref{int}) and substituting there
$\rho_i=\lambda_it$ (strictly speaking, one had to average the
logarithm over the initial measure as well, yet this differs by a
correction decreasing as $t^{-1}$). At $t\ll t_d$ the growth rate
$\gamma=\lambda_1$, as it has to be for a perfect conductor.  During
an intermediate stage $t\sim t_d$ the growth rate decreases and
eventually at $t\gg t_d$ it comes to an asymptotic value
$\gamma_{\infty}$, independent of $\kappa$ (so called fast dynamo
\cite {CG95,VZ72}):
\begin{equation}\gamma_{\infty}=\min\left\{
{(\lambda_1-\lambda_2)}/{2},\,(\lambda_2-\lambda_3)/{2}\right\}.
\label{gamma}
\end{equation}
Note that the growth-rate does not fluctuate at large times,
$\gamma \geq0$ and $\gamma_\infty\to0$ as
$\lambda_2\to\lambda_1$ or $\lambda_2\to\lambda_3$, corresponding to
the zero growth rate for axially symmetric cases. For a
time-reversible flow statistics $\lambda_2=0$ and
$\gamma_\infty=\lambda_1/2$.  For isotropic Navier-Stokes turbulence,
numerical data suggest $\lambda_2\simeq\lambda_1/4$ \cite{GP90}, so
that our prediction for the long-time growth rate in a typical realization is
$\gamma_\infty\simeq3\lambda_1/8$.

The moments with $n>0$ all grow in a random incompressible flow with a
nonzero Lyapunov exponent since $E_n=\log \langle B^{2n}\rangle /2t$
is a convex function of $n$ (due to H\"older inequality) with $E_0=0$
and $dE_n/dn(0)=\gamma \geq0$.  Even when $\gamma=0$, the growth rates
$E_n$ are positive for $n>0$ if $H$ has a finite width, that is if the
flow is random (for $n=1$ this was previously stated in \cite{GCS96}).
The growth of the $2n$-th moment at $t\ll t_d$ is determined by the
average of $\exp(2n\rho_1)$. For $t\gg t_d$, the expression to average
is either $\exp(n\rho_1-n\rho_2)$ (with $\rho_2> \log r_d/L$) or
$\exp(n\rho_2-n\rho_3)$ depending on whether the entropy function
favours positive or negative $\rho_2$ (cf. \cite{MRS85}).  The
function $E_n$ is nonuniversal since it is determined by the
saddle-point of ${\cal D}\rho_{1,2}$ integration which depends on the
particular form of the entropy function. The saddle-point falls within
the (universal) parabolic region of $H$ around the minimum only for
$n\ll(\lambda_1\tau)^{-1}$. Therefore, the temporal growth of the
moments will be calculated below for a short-correlated velocity.

Here, we continue with the general case to establish what is universal
in the different-point correlation functions $F_{2n}=\langle
\prod_{k=1}^n\left({\bf B}({\bf x}_{2k-1},t){\bf B}({\bf
x}_{2k},t)\right) \rangle$. Its calculation is reduced to averaging
$(2n-1)!!$ terms arising from the Wick decomposition in the Gaussian
integration over the random initial condition. Each term is a product
of $n$ integrals generalizing the one for $B^2$ with the inclusion of
respective $\exp(i{\bf r}_j \hat{W}^{T,-1}{\bf q})$ in the
integrand. The $n$ vectors ${\bf r}_j$ are differences between couples
of ${\bf x}_k$'s.  The new feature with respect to the moments is the
presence of the rotation matrix $\hat{R}$ in the exponential factor.
The ${\bf q}$-integrations proceed along the same lines as previously:
in any of the $n$ integrals we change variables ${\bf q}=\hat{S}^T{\bf
Q}$ and the dependence on $\hat{S}$ is thus entirely moved into the
prefactors. Substituting $\hat{S}$ by the unit matrix and performing
the ${\bf Q}$-integrations, we obtain the long-time asymptotics for
any of the $(2n-1)!!$ contributions to $F_{2n}$\,:
\begin{eqnarray} &&
\Biggl\langle \frac{({L}/{r_d})^n\exp\left[n(\rho_1-\rho_2)\right]}{
\left[1+e^{-2\rho_2}r_d^2/L^2\right]^{5n/2}}\int_{-1}^1\! d\cos\theta
\int_0^{2\pi}\!d\varphi \int_0^{2\pi}\!
 d\phi
\label{F2n}\\ &&\times\prod_{j=1}^n
\left[2-\frac{R_{2j}^2e^{-2\rho_2}}{L^2}\right]
\exp\left[-\frac{R_{2j}^2}{4\left[L^2e^{2\rho_2}+2r_d^2\right]}
-\frac{R_{3j}^2}{8r_d^2}\right]
\Biggr\rangle\nonumber
\end{eqnarray}
where ${\bf
R}_j=\hat{R}_3[\varphi]\hat{R}_2[\theta]\hat{R}_3[\phi]{\bf r}_j$, and
$\hat{R}_{2,3}$ stand for rotation around $Y$ and $Z$ axis
respectively.  We consider time-reversible statistics, where the
scaling laws turn out to be universal. Let us explain the physical
meaning of (\ref{F2n}) and derive the correlation functions starting
from $n=1$.  The realizations contributing have the advective exponent
$\exp(i{\bf r} \hat{W}^{T,-1}{\bf q})$ of order unity.  This requires
$\rho_2>\ln(r/L)$ and the direction of contraction ${\bf n}_3$ to be
almost perpendicular to ${\bf r}$, which gives angular factor
$(L/r)\exp\rho_3$. At $\lambda_1t>\ln(L/r)$ we get then
$$F_2(r,t)\simeq{L\over r}\int_{-\infty}^\infty\! d\rho_1
\int_{\ln(r/L)}^\infty \!d\rho_2e^{\rho_1-\rho_2-tH}\propto
r^{-2-h}e^{E_2t},$$ where $h$$=\partial H/\partial y$ taken at $y=0$
and at $x$ given by the saddle point $\partial H/\partial x=1$. Time
reversibility means that $H(x,y)=H(x+y,-y)$ so $h=1/2$ and $F_2\propto
r^{-5/2}\exp(E_2t)$. At $r\ll L$ and $\lambda_1t$$\ll \ln(L/r)$, $F_2$
is $r$-independent. This generalizes the consideration of \cite{KA92}
for an arbitrary time-reversible statistics.  To understand the simple
geometrical picture behind this derivation, note that the integral
over $\rho_2$ comes from $\rho_2\simeq\ln(r/L)$. That means that the
field configurations in the form of strips with width $r$ dominate
$F_2(r)$. Angular integral in (\ref{F2n}) comes from $\varphi\simeq1$
that is the strips with the stretching direction almost parallel to
${\bf r}$ do not contribute (because of cancellations due to
solenoidality).

For $n\geq2$, the geometry of the vectors ${\bf r}_j$ becomes
important.  Let us first consider the case where all the vectors are
in the same plane (their length being $r$).  They can be either
collinear or not. Almost orthogonality of ${\bf r}_j$ to ${\bf n}_3$
involves therefore either one angle or two, giving the angular factor
$(L/r)\exp\rho_3$ or its square, respectively. The other difference
concerns the behavior along ${\bf n}_2$.  For non-collinear geometry
not all vectors can be orthogonal to ${\bf n}_2$ and $\rho_2$ should
then be constrained as $\rho_2>\log r/L$. This is technically
signalled by the fact that the integration over $\varphi$ is not
saddle-point. Conversely, for collinear geometry all the vectors can
be orthogonal both to ${\bf n}_2$ and ${\bf n}_3$, giving an
additional angular factor $(L/r)\exp\rho_2$\,: the saddle-point
integrations over $\theta$ and $\varphi$ pick $\theta=\pi/2$,
$\varphi\simeq \exp[\rho_2]L/r$. In the rest of the integrals (either
$n-1$ or $n-2$, respectively) the above angular constraints ensure
that the advective exponents are $O(1)$ so the diffusive exponents
$\exp(-2\kappa{\bf q}\Lambda{\bf q})$ become important.  The
calculation of these integrals is essentially the same as for the
moments and this is where diffusion comes into play. The wavevectors
should be quasi orthogonal to ${\bf n}_3$, giving either $n-1$ or
$n-2$ factors $(L/r_d)\exp\rho_3$. The growth along ${\bf n}_2$ for a
generic planar geometry is automatically controlled by the previous
constraint $\rho_2>\log r/L$\,; for collinear geometry it provides the
bound $\rho_2>\log r_d/L$. Simply speaking, the strips with the width
$r_d$ stretched along ${\bf r}$ contribute in the collinear case,
while the width is $r$ in the generic case. The resulting integrations
over $\rho_1$ are saddle-point and those over $\rho_2$ are dominated
by the lower bounds. Finally
\begin{equation}
F_{2n}\simeq e^{E_n t}\left(\frac{L}{r_d}\right)^{{5n\over 2}}
\left(\frac{r_d}{r}\right)^2\times
\left\{
\begin{array}{cc}
1 & \mbox{collinear}\\
(r_d/r)^{{3n\over 2}} & \mbox{planar}.
\end{array}
\right.
\end{equation}
Here, $E_n=x_n-H(x_n,0)$ with ${\partial H/\partial x}(x_n,0)=n$. The
fact that the integrals over $\rho_2$ are all dominated by the lower
bounds indicates the geometric nature of the scaling universality
found\,: the magnetic field configurations that contribute are narrow
strips (not ropes and layers suggested in \cite{MRS85}) with one
direction of stretching, one of contraction and a neutral one. The
factor $(r_d/r)^2$ is the probability for two points at distance $r$
to lie within the same strip of width $r_d$.

The peculiar nature of strip configurations has another dramatic
consequence for $n\ge 3$: the correlation functions are strongly
suppressed in a generic situation when at least three vectors ${\bf
r}_j$'s do not lie in parallel planes. Indeed, they cannot then be on
parallel strips and nonzero correlation appears only because the
strips have exponential diffusive tails.  As a result, the factor
$(r_d/r)^{2}$ in the planar formula is replaced by $\exp(-a{r^2}
\sin^2\Theta/{r_d^2})$, where $a\simeq1$ and $\Theta$ is the minimal
angle between a vector and the plane formed by another two.  That can
be derived from (\ref{F2n}) where all angular integrations are not
saddle-point now. For a general irreversible velocity statistics, 
the $r$-dependencies are  different 
yet the qualitative conclusions (that the
correlation functions are not exponentially suppressed only for planar geometry
and are anomalously large for collinear geometry) are generally valid.
Note that angular anomalies are likely to be
peculiar of the viscous interval where advection by a smooth velocity
preserves collinearity. Similar collinear anomalies 
 (which are 
manifestations of the strip configurations described here)
have been found before for a passive scalar
advected by a smooth velocity \cite{BCKL95,BFLL99}. Note
that the cliff-and-ramp structures observed in passive scalar experiments
(see \cite{91Sre} for review) are probably related to the strips too.

It is left to find $E_n$ for the standard Kazantsev-Kraichnan model of
an isotropic short-correlated Gaussian strain with 
$\langle\sigma_{\alpha\beta}(t)\sigma_{\alpha\beta}(0)\rangle=10\lambda_1
\delta(t)$.  Straightforward 
derivation gives Gaussian ${\cal
P}_t(\rho_1,\rho_2)$ with 
$H={(\rho_1+\rho_2/ 2- \lambda_1
t)^2/ \lambda_1t}+{3\rho_2^2/ 4\lambda_1t}$. 
Note that $\lambda_2=0$. 
Now we integrate the moments of
(\ref{int}) with such ${\cal
P}_t$.  First, we integrate $ \exp(2n\rho_1)$ and 
get the answer for the perfect conductor 
\cite{CG95,MRS85}
\begin{equation}
\langle {\bf B}^{2n}\rangle$$ \simeq \exp \left[ 2{\lambda_1}n(2n+3)t/3\right]\ .
\label{moments}
\end{equation}
The main contribution
to the $n$-th moment comes from 
$\rho_1=\lambda_1t(4n+3)/3,\rho_2=-2n\lambda_1t/3$ so (\ref{moments}) is valid 
until $L\exp\rho_3>r_d$ that is for
 $t<3t_d/(2n+3)$ with $t_d\equiv\lambda_1^{-1}\ln(L/r_d)$. 
Then there is a logarithmically wide
cross-over interval when 
the growth is non-exponential. The asymptotic regime starts at
$t>3t_d/(n+2)$ when 
unity in the first parenthesis of (\ref{int}) may be neglected. 
The integral over $\rho_1$
now  comes from $\rho_1+\rho_2/2=(n+2)\lambda_1t/2$
while that over $\rho_2$ is dominated by 
$\rho_2\simeq\ln(r_d/L)$:
\begin{equation}
\langle {\bf B}^{2n}\rangle \simeq\left({L/ r_d}\right)^{5n/2}\exp(E_nt),\ 
E_n=\lambda_1n(n+4)t/4\,.\label{momenty}\end{equation} For $n=1$, this
was obtained by Kazantsev \cite{K68}. 
The difference between
(\ref{moments}) and (\ref{momenty}) formally means that the two limits
$t\to\infty$ and $\kappa\to0$ do not commute (what is called
dissipative anomaly).  The physical reason is quite clear\,:
realizations with continuing contraction along two directions
contribute most in a perfect conductor, while with diffusion present
one direction is neutral. 
A magnetic field initially concentrated in
the ball with the radius $L$ will have the fastest growth rate $\gamma$
if the ball turns into a
strip with the dimensions 
$L(L/r_d)^{1/2}\exp(3Dt)$, $r_d$, $L(r_d/L)^{1/2}\exp(-3Dt)$. 
The main contribution into the $n$-th moment will be given by strips with the
dimensions $L(L/r_d)^{1/2}\exp[3(n+2)Dt/2]$, $r_d$ and
$L(r_d/L)^{1/2}\exp[-3(n+2)Dt/2]$. To conclude,
we related the growth-rate of the small-scale dynamo 
to the Lypunov exponents of the flow and described analytically
the strip structure of magnetic field.

Very fruitful discussions with E. Balkovsky, A. Fouxon, U. Frisch,
A. Gruzinov, V. Lebedev, A. Pouquet, B. Shraiman and A. Vulpiani are
gratefully acknowledged.  The work was supported by a JRO fellowship
(MC), by the Einstein Center and the grant of Minerva Foundation at
the Weizmann Institute (GF, IK) and by the Russian Foundation for
Basic Research (IK).

\end{multicols}


\begin{references}
\bibitem{M78}H.K.~Moffatt,  {\it Magnetic field generation in electrically
conducting fluids} (Cambridge Univ. Press, 1978).

\bibitem{P79}  E.N.~Parker, {\it Cosmic magnetic fields, their origin and
activity} (Clarendon Press, Oxford, 1979).

\bibitem{CG95} S.~Childress and A.~Gilbert, {\it Stretch, Twist, Fold:
The Fast Dynamo} (Springer-Verlag, Berlin, 1995).

\bibitem{B50} G.K.~Batchelor, Proc. Roy. Soc. London A {\bf 261}, 405
(1950).

\bibitem{KN67} R.~Kraichnan and S.~Nagarajan, Phys. Fluids {\bf 10},
859 (1967).

\bibitem{K68}A.P.~Kazantsev,  
Sov. Phys JETP {\bf 26}, 1031 (1968).

\bibitem{KA92} R.~Kulsrud and S.~Anderson, 
Astrophys. J. {\bf 396}, 606 (1992).

\bibitem{VZ72}  S.~Vainshtein and Ya.~Zeldovich,  
Sov. Phys. Usp. {\bf 15}, 159 (1972).

\bibitem{GP90} S.~Girimaji and S.~Pope,
J. Fluid Mech. {\bf220}, 427(1990).

\bibitem{MRS85}  S.~Molchanov, A.~Ruzmaikin, and A.~Sokolov, 
Sov. Phys. Usp. {\bf28}, 307--327 (1985).

\bibitem{94SS}  B.~Shraiman and E.~Siggia, 
Phys.~Rev. E {\bf 49}, 2912 (1994).

\bibitem{95CFKLa}  M.~Chertkov, G.~Falkovich, I.~Kolokolov and V.~Lebedev,
Phys. Rev. E {\bf 51}, 5609 (1995).

\bibitem{59Bat} G.K.~Batchelor, J. Fluid. Mech. {\bf 5}, 113 (1959).

\bibitem{74Kra} R.H.~Kraichnan, J. Fluid. Mech. {\bf 64}, 737 (1974).

\bibitem{CPV} A.~Crisanti, G.~Paladin and A.~Vulpiani, 
{\it Products of random matrices} (Springer-Verlag, Berlin 1993). 

\bibitem{GSO87} I.~Goldhirsch, P.-L.~Sulem and S.~Orszag,  
Physica D {\bf27}, 311 (1987).

\bibitem{GCS96} A. Gruzinov, S. Cowley and  R. Sudan, 
Phys. Rev. Lett. {\bf 77}, 4342--4345 (1996).

\bibitem{Ellis} R. Ellis, {\it Entropy, Large Deviations and Statistical
Mechanics} (Springer Verlag 1985).

\bibitem{BF99} E. Balkovsky and A. Fouxon, chao-dyn/9905020.

\bibitem{BCKL95}
E. Balkovsky, M. Chertkov, I. Kolokolov and V. Lebedev,
JETP Lett. {\bf 61},
1049 (1995).

\bibitem{BFLL99} E. Balkovsky, G. Falkovich, V. Lebedev and 
M. Lysiansky, Phys. Fluids (in press).

\bibitem{91Sre} K.R.~Sreenivasan, 
Proc.R. Soc. Lond. {\bf A} 434, 165 (1991).
\end{references}
\end{document}